\documentclass[useAMS,usenatbib,usegraphicx]{mn2e}
\usepackage{times}

\newif\ifAMStwofonts
\AMStwofontstrue

\newcommand{\simlt}{\lower.5ex\hbox{$\; \buildrel < \over \sim \;$}}
\newcommand{\simgt}{\lower.5ex\hbox{$\; \buildrel > \over \sim \;$}}
\newcommand{\be}{\begin{equation}}
\newcommand{\ba}{\begin{eqnarray}}
\newcommand{\ee}{\end{equation}}
\newcommand{\ea}{\end{eqnarray}}

\def\Sigstel{\Sigma_{\rm stel}} \def\Sigtot{\Sigma_{\rm tot}}
\def\Mstel{M_{\rm stel}} \def\Mtot{M_{\rm tot}}
\def\athird{{\textstyle\frac13}}
\def\Re{R_e} \def\Rm{R_M}

\title[Mass maps of gravitationally lensing galaxies]
      {Unveiling dark halos in lensing galaxies}
\author[I. Ferreras, P. Saha and S. Burles]
{Ignacio Ferreras$^1$\thanks{Current address: Mullard Space Science Laboratory/UCL,
Holmbury St. Mary, Dorking, Surrey RH5 6NT}, 
Prasenjit Saha$^2$ and Scott Burles$^3$.\\
$^1$ Dept. of Physics, King's College London, Strand, London WC2R 2LS\\ 
$^2$ Institute for Theoretical Physics, University of Z\"urich,
  Winterthurerstrasse 190,8057 Z\"urich, Switzerland\\
$^3$ Dept. of Physics, Massachusetts Institute of Technology,
  77 Massachusetts Ave., Cambridge, MA 02139, USA
}
\begin{document}
\date{Accepted for publication in MNRAS 2007 October 16.}
\pagerange{\pageref{firstpage}--\pageref{lastpage}} \pubyear{2007}
\maketitle
\label{firstpage}

\begin{abstract}
We present a spatially resolved comparison of the stellar-mass and
total-mass surface distributions of nine early-type galaxies. The
galaxies are a subset of the Sloan Lens ACS survey \citep[or
SLACS;][]{2006ApJ...638..703B}.
% redshifts from 0.081 to 0.322.
The total-mass distributions are obtained by exploring pixelated mass
models that reproduce the lensed images.  The stellar-mass
distributions are derived from population-synthesis models fit to the
photometry of the lensing galaxies.  Uncertainties -- mainly
model degeneracies -- are also computed.  Stars can account for all the
mass in the inner regions.  A Salpeter IMF actually gives too much
stellar mass in the inner regions and hence appears ruled out.  Dark
matter becomes significant by the half-light radius and becomes
increasingly dominant at larger radii.  The stellar and dark
components are closely aligned, but the actual ellipticities are not
correlated.  Finally, we attempt to intuitively summarize the results
by rendering the density, stellar-vs-dark ratio, and uncertainties
as false-colour maps.
\end{abstract}


\section{Introduction}

In the current paradigm of galaxy formation, the building block of
structure is a dark matter halo, consisting mainly of non-baryonic
dark matter together with $\sim 15\%$ baryons in the form of gas.  Dark
halos are thought to originate from the collapse of primordial density
fluctuations, growing unimpeded until virial equilibrium is
reached. Within these halos the baryonic component dissipates energy,
collapsing further towards the center and eventually forming the visible
galaxy. Subsequent mergers redistribute the matter within halos.

To work out the details of the basic picture, which goes back to
\cite{1978MNRAS.183..341W}, it is essential to determine the
connection between the visible galaxies and dark halos. Strong
gravitational lensing by galaxies is potentially a very useful way of
doing this, since the total mass of a lensing galaxy is relatively
easy to constrain.  Also, since lensing tends to be more effective for
distant galaxies ($z_{\rm lens}\sim0.1$ to 1), it nicely complements
the stellar-dynamical techniques applicable in nearby galaxies.

The difficulty with lensing galaxies (that is to say, galaxies
producing multiple images of background sources) is that they are
relatively rare.  Till recently, only about 80 were known.  But
recently, 28 new galaxy lenses have been discovered by the Sloan Lens
ACS Survey \citep[SLACS;][]{2006ApJ...638..703B} thanks to a new survey
strategy, which eventually may double the number of strong lenses.
The method is to select galaxies from the Sloan survey
\citep{2000AJ....120.1579Y} that have emission lines indicating
high-redshift background objects, and then to image these candidates
using the Advanced Camera for Surveys on the Hubble Space Telescope.

A basic analysis of a sample of galaxy lenses is to fit simple lens
models and then compute $M/L$ and its evolution with redshift.  This
was first done by \cite{1998ApJ...509..561K}.  An extension is to
place lensing galaxies on the Fundamental Plane of ellipticals, using
a measured or model-derived velocity dispersion
\citep{2000ApJ...543..131K,2003ApJ...587..143R,2006ApJ...640..662T}.
Other work \citep{2004ApJ...611..739T,2006ApJ...649..599K}
compares lens models with the measured dispersions to
constrain the mass profiles of the galaxies.  These studies have found
no unexpected features or trends with redshift, and argue in favor of
passive evolution.

A more detailed analysis involves modeling both the lens mass
distribution and the stellar population. The star-formation
history is not well-constrained by the observed fluxes and colours and
must be marginalized over, but the stellar mass is fairly insensitive
to model assumptions apart from the initial mass function (IMF).  The
lensing mass distribution, when aggressively modeled, turns out to
have much larger uncertainties than simple models assume;
nevertheless, the uncertainties can be estimated and useful
conclusions drawn.  \cite{2005ApJ...623L...5F} found massive
ellipticals to show a transition from no significant dark matter
within $\Re$ to dark-matter dominance by $5\Re$, whereas lower-mass
galaxies showed no significant dark halos even at $5\Re$. The radial
gradient in the dark-matter fraction agreed with the results on nearby
galaxies derived from stellar dynamics \citep{2005MNRAS.357..691N}.

In this paper we extend the detailed comparison of stellar and total
mass to two dimensions, using a subsample of the SLACS lenses.  All
the SLACS objects have a background galaxy that is lensed into two or
four extended images. In nine of the objects, we can identify small
features within the extended images.  For lenses showing point-like
multiply-imaged features there is a well-developed technique for
reconstructing the projected mass distribution, along with uncertainty
estimates \citep{2004AJ....127.2604S}.  Accordingly we take these
nine lenses as our sample.  The stellar mass content is estimated by
combining the available photometry of the lensing galaxies with
population-synthesis models \citep{2003MNRAS.344.1000B}.

%%%%%%%%%%%%%%%%%%%%%%%%%%%%%%%%%%%%%%%%%%%%%%%
%%%%%%%%%%%%     FIGURE 1   %%%%%%%%%%%%%%%%%%%
%%%%%%%%%%%%%%%%%%%%%%%%%%%%%%%%%%%%%%%%%%%%%%%
\begin{figure}
\begin{center}
\includegraphics[width=3.4in]{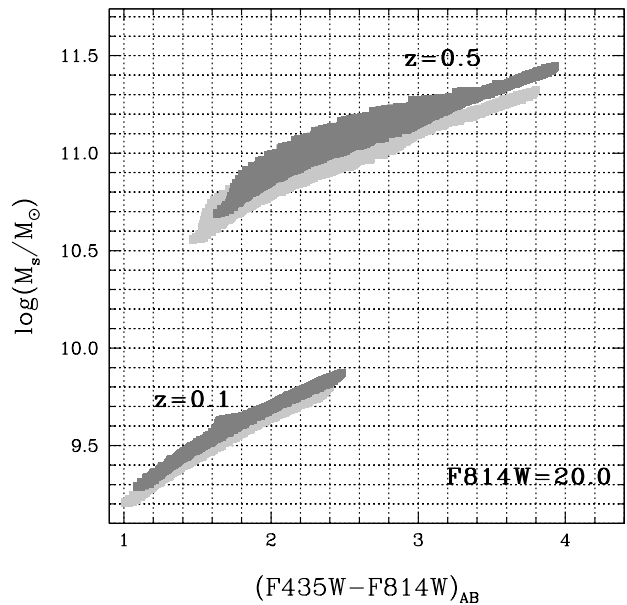}
\caption{Stellar masses derived from the available photometry. The
predictions of the Bruzual \& Charlot (2003) models are shown for a
Chabrier (2003) IMF. A galaxy with apparent magnitude 
$\rm F814W=20$ is considered at two different redshifts as labelled. The
shaded regions span the model predictions when assuming a wide range
of $\tau$ models (i.e., an exponentially decaying star formation
history at fixed metallicity). The dark/light shaded regions
correspond to a formation redshift of $z_F=5$ and $2$, respectively,
and all span a range of star formation timescales 
($-1<\log \tau ({\rm Gyr}) < +1$) and metallicities ($-1<\log Z/Z_\odot <+0.3$).}
\label{fig:stellar}
\end{center}
\end{figure}
%%%%%%%%%%%%%%%%%%%%%%%%%%%%%%%%%%%%%%%%%%%%%%%

\section{Mapping the total mass}\label{sec:totmass}

We use the pixelated lens reconstruction method implemented in the
{\em PixeLens\/} program.\footnote{Available from {\tt
http://www.qgd.uzh.ch/projects/pixelens/}} The algorithm is motivated
and described in detail in \cite{2004AJ....127.2604S}, but basically
it consists of two ideas.

The first idea is to express the lensing data as a set of linear
constraint equations on the mass distribution.  We assume that the
centroids of multiple images are measured with negligible error, which
--- at HST resolution --- is a very good approximation.  Then the
remaining unknowns are the source position and the lens mass
distribution, and both of these enter the lens equation linearly
\citep[cf.~Equation 5 in][]{1997MNRAS.292..148S}.  The lens is
expressed as a superposition of mass tiles or pixels, although a
superposition of basis functions is also possible
\citep{2000ApJ...535..671T}.  A prior on the lens can be expressed as
linear inequalities.  Specifically, we require the mass distribution
to (i)~be non-negative, (ii)~be centrally concentrated, with the local
density gradient pointing $\leq 45^\circ$ away from the center,
(iii)~be inversion symmetric (i.e. symmetric with respect to rotation by
$180^\circ$; optional),
(iv)~have no pixel more than twice the sum of its neighbors, except
possibly the central pixel, and (v)~the circularly averaged mass
profile to be steeper than $R^{-0.5}$, where $R$ is the projected
radius.  Item (v) is assumed (for galaxies) since stellar dynamics or
other methods never give $\rho(r)$ as shallow as $r^{-1.5}$
\citep[e.g.,][]{1991MNRAS.252..210B,2001AJ....121.1936G}.  These
various equations and inequalities give an underdetermined linear
system analogous to Schwarzschild's problem in stellar dynamics.

The second idea \citep[suggested by][]{2000AJ....119..439W} is to
sample the mass maps allowed by the data and prior, through a
Monte-Carlo method.  The result is an ensemble of lens
models. Uncertainties on any parameter of the lens can be derived from
the model ensemble in the usual way. The ensemble-average will
automatically satisfy the data and prior constraints, since they are
linear.  This makes the ensemble-average model a good choice for
representing a single model. For a detailed illustration of the
ensemble method, see the Appendix in \cite{mesostrucSaha}.

%%%%%%%%%%%%%%%%%%%%%%%%%%%%%%%%%%%%%%%%%%%%%%%
%%%%%%%%%%%%     FIGURE 2   %%%%%%%%%%%%%%%%%%%
%%%%%%%%%%%%%%%%%%%%%%%%%%%%%%%%%%%%%%%%%%%%%%%
\begin{figure}
\begin{center}
\includegraphics[width=3.4in]{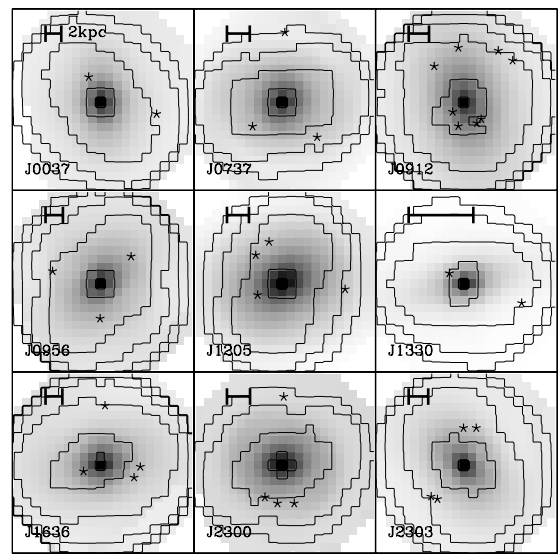}
\caption{Projected mass maps of the lensing galaxies. The gray scale
represents $\Sigstel$ while the contours show $\Sigtot$.  The latter
are in multiplicative steps of $10^{0.4}$ (like a magnitude scale).
The star symbols mark multiply-imaged features of the background
source. The bar in each panel indicates 2~kpc.}
\label{fig:mass}
\end{center}
\end{figure}
%%%%%%%%%%%%%%%%%%%%%%%%%%%%%%%%%%%%%%%%%%%%%%%

For each lens we computed an ensemble of 200 mass maps of $21\times21$
pixels each.  To derive radial profiles we circularly-averaged the
mass maps.  This procedure naturally yields ensembles of radial
profiles, hence an estimate of the uncertainty.

In the lens models the enclosed mass is best constrained at projected
radii similar to the images themselves (roughly speaking, around the
Einstein radius), whereas at smaller and larger radii, the enclosed
mass becomes progressively more uncertain.  This is simply the
well-known steepness degeneracy in lensing theory \citep[see
e.g.,][]{2000AJ....120.1654S}.  As a result, the enclosed mass
profile and its error-bars have a characteristic butterfly shape.

To what extent velocity dispersions can break lensing degeneracies
remains an open question.  Studies using velocity dispersions as
constraints on lenses tend to be optimistic.  However,
stellar-dynamical degeneracies associated with kinematics are also
known \citep{2004ApJ...602...66V,2004MNRAS.347L..31C} and it is not
clear that these will be orthogonal to lensing degeneracies.  Also,
current methods for incorporating kinematics into lens models assume
spherical symmetry.  For these reasons we do not include kinematics in
the lens models.

We do not attempt to fit the extended images.  Now, a model fitted to
point-like images automatically provide a predictions for arcs that
would be generated by a conical light profile
(i.e. a circularly symmetric model; \cite{2001AJ....122..585S})
and these predicted arcs generally do yield a good approximation to
observed arcs.  Hence disregarding the extended images does not
sacrifice as much information as might at first appear.  Fitting a
better source distribution is a straightforward linear
image-reconstruction problem if the lens model is fixed.  But using
the source structure to improve the lens fit is much harder.  Several
papers propose schemes that iteratively fit lens and source models.
But they either assume a restricted parameterized form for the lens
\citep[e.g.,][]{1994ApJ...426...60W,2005ApJ...623...31D,2006ApJ...651....8B,2006ApJ...649..599K}
or allow the lens model to be free form
\citep{2005MNRAS.363.1136K,2006MNRAS.366...39S} but need it to be
close to a starting guess.

%%%%%%%%%%%%%%%%%%%%%%%%%%%%%%%%%%%%%%%%%%%%%%%
%%%%%%%%%%%%     FIGURE 3   %%%%%%%%%%%%%%%%%%%
%%%%%%%%%%%%%%%%%%%%%%%%%%%%%%%%%%%%%%%%%%%%%%%
\begin{figure}
\begin{center}
\includegraphics[width=3in]{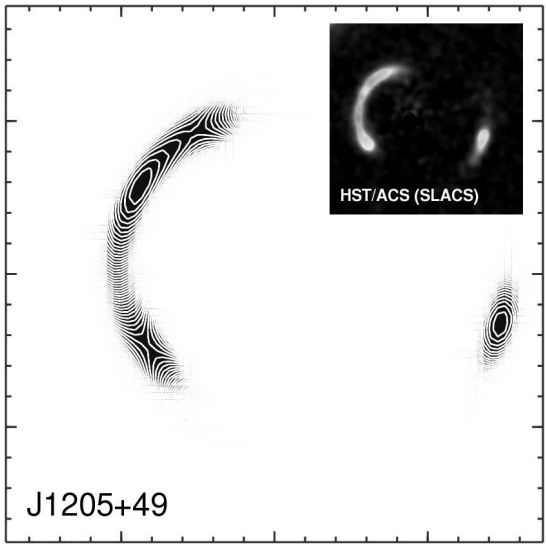}
\caption{Extended images in the lens J1205+491. {\em Main
panel.}~Extended images predicted by our model for this lens, for a
circular source of size $0.15''$ having a conical light profile. {\em
Inset.}~The observed image with the lensing galaxy substracted off.}
\label{fig:J1205}
\end{center}
\end{figure}
%%%%%%%%%%%%%%%%%%%%%%%%%%%%%%%%%%%%%%%%%%%%%%%

\section{Mapping the stellar mass}

The galaxy images are Wide-Field Channel/ACS snapshots with an
exposure time of 420s in each of the F435W ($B$) and F814W ($I$)
filters.  For each system we worked on the best-fit smooth lens-galaxy
image that was originally subtracted from the actual image in order to
discern faint lensing features. The lens-galaxy images were further
corrected for Galactic reddening using the available dust maps
\citep{1998ApJ...500..525S} and a standard model for extinction by
Galactic dust \citep{1999PASP..111...63F}.  We then spatially rebinned
the photometry of each galaxy into the same $21\times21$ pixelation
used for the lens models of that galaxy.  These pixels are $\sim0.1''$
to $0.2''$ across, depending on the system, amounting to $\sim0.1$ to
1\thinspace kpc at the lensing galaxy.

For the stellar-population analysis we apply the so-called $\tau$
models, which assume an exponentially decaying star formation history
at fixed metallicity.  Three parameters describe each model: the
formation epoch (i.e., when star-formation starts), the formation
timescale and the metallicity.  We explore a grid of $64\times
64\times 64$ models and allow in our estimates only those models whose
photometry is compatible with the observations, treating the 
$21\times21$ pixels described above as independent.  One colour is not
sufficient for an estimate of the stellar ages and metallicities, but
is enough to constrain the stellar mass content, because of the wide
spectral range covered by $B-I$.  Figure~\ref{fig:stellar} illustrates
this point. In this figure we consider a \cite{2003PASP..115..763C}
Initial Mass Function with formation epoch $z_F=2$ or $z_F=5$, viewed
from $z=0.1$ or $0.5$. The stellar mass corresponds to a galaxy with
apparent magnitude $\rm F814W=20$ at the labelled redshifts.  The
shaded regions show the range of $B-I$ as the metallicity varies in
the range $-1<\log Z/Z_\odot <+0.3$ and the star formation timescale
varies as $-1<\log \tau ({\rm Gyr}) < +1$. The uncertainty in ages and
metallicities result in a $\sim 0.2$--$0.3$~dex uncertainty in the stellar
mass content.

%%%%%%%%%%%%%%%%%%%%%%%%%%%%%%%%%%%%%%%%%%%%%%%
%%%%%%%%%%%%     FIGURE 4   %%%%%%%%%%%%%%%%%%%
%%%%%%%%%%%%%%%%%%%%%%%%%%%%%%%%%%%%%%%%%%%%%%%
\begin{figure}
\begin{center}
\includegraphics[width=3.4in]{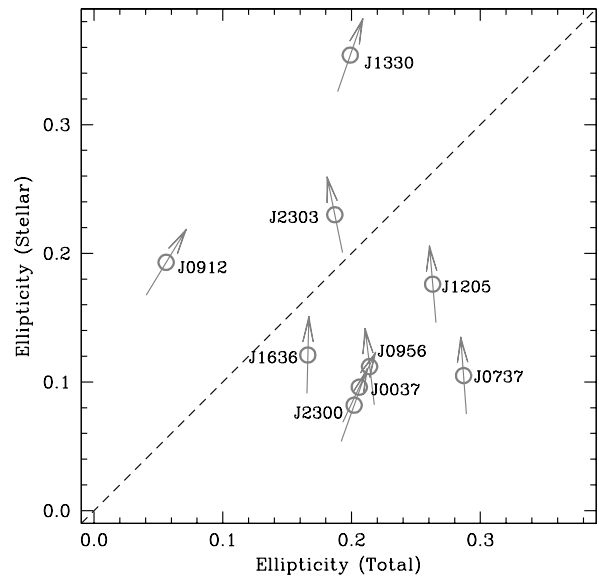}
\caption{A comparison of the ellipticity of the total and stellar mass
distributions. The dots give the values of the measured ellipticities
whereas the orientation of the arrows represent the misalignment
between both distributions; a vertical arrow means perfectly aligned.}
\label{fig:ellip}
\end{center}
\end{figure}
%%%%%%%%%%%%%%%%%%%%%%%%%%%%%%%%%%%%%%%%%%%%%%%

Our sample comprises early-type galaxies. In these systems the
presence of an overall old and coeval stellar population
\citep{1998ApJ...492..461S} suggests star formation took place in a
strong burst that consumed most of the available gas
\citep{2003MNRAS.344..455F}. Other baryonic components, hot and cold
gas or dust, are known to contribute a small fraction to the net
baryon budget in these systems \citep{1994ARA&A..32..115R}. Hence, we
can take the stellar mass content in each modeling pixel as the baryon
content.

%%%%%%%%%%%%%%%%%%%%%%%%%%%%%%%%%%%%%%%%%%
%%%%%%%%%%%%%%%%  Table  %%%%%%%%%%%%%%%%%
%%%%%%%%%%%%%%%%%%%%%%%%%%%%%%%%%%%%%%%%%%
\begin{table*}
\begin{minipage}{13cm}
\caption{Summary of the galaxy sample. (Indicated errors are 90\%
confidence.)\label{tab:sample}}
\begin{tabular}{lrcccrrr} \hline\hline
ID & $\Re^1$ & Redshift$^1$ & $\sigma^1$ & $R_{\rm M}$ &
M$_{\rm TOT}(<R_{\rm M})$ & M$_{\rm STAR}(<R_{\rm M})$ &
$5\sigma^2\Re /G$\\
    & kpc  &    & km/s  &  $/\Re$ &  $10^{10}M_\odot$ & 
 $10^{10}M_\odot$ &  $10^{10}M_\odot$\\
\hline
J003753.21-094220.1  &  $6.99$ & $0.195$ & $265$ & $1.66$ & $52.3_{-14.5}^{+35.4}$ & $44.3_{-4.3}^{+5.1}$ & $57.1$\\
J073728.45+321618.5  & $10.10$ & $0.322$ & $310$ & $0.79$ & $38.3_{ -3.9}^{+11.8}$ & $40.2_{-3.6}^{+4.1}$ & $112.9$\\
J091205.30+002901.1  &  $9.47$ & $0.164$ & $313$ & $0.95$ & $71.2_{-21.7}^{+14.5}$ & $44.7_{-4.0}^{+4.6}$ & $107.9$\\
J095629.77+510006.6  &  $8.85$ & $0.241$ & $299$ & $1.20$ & $66.4_{-16.7}^{+25.7}$ & $41.8_{-4.0}^{+4.9}$ & $92.0$\\
J120540.43+491029.3  &  $8.03$ & $0.215$ & $235$ & $1.04$ & $38.6_{ -5.5}^{+16.6}$ & $50.5_{-4.6}^{+5.1}$ & $51.6$\\
J133045.53-014841.6  &  $1.28$ & $0.081$ & $178$ & $2.19$ &  $4.9_{ -1.7}^{ +3.5}$ &  $3.8_{-0.4}^{+0.4}$ & $4.7$\\
J163602.61+470729.5  &  $5.41$ & $0.228$ & $221$ & $2.01$ & $52.9_{-14.3}^{+16.6}$ & $32.2_{-3.2}^{+3.8}$ & $30.7$\\
J230053.14+002237.9  &  $6.43$ & $0.229$ & $283$ & $1.24$ & $40.9_{ -3.1}^{+14.2}$ & $28.0_{-2.6}^{+3.0}$ & $59.9$\\
J230321.72+142217.9  &  $8.13$ & $0.155$ & $260$ & $1.17$ & $49.8_{-15.5}^{+22.8}$ & $37.5_{-3.5}^{+4.0}$ & $63.9$\\
\hline\hline
\end{tabular}
%\tablenotetext{1}{Data from Bolton et al. (2006).}
\medskip
$^1$ Data from Bolton et al. (2006).
\end{minipage}
\end{table*}
%%%%%%%%%%%%%%%%%%%%%%%%%%%%%%%%%%%%%%%%%

\section{Comparison of stellar and total mass}

\begin{figure}
\begin{center}
\includegraphics[width=3.4in]{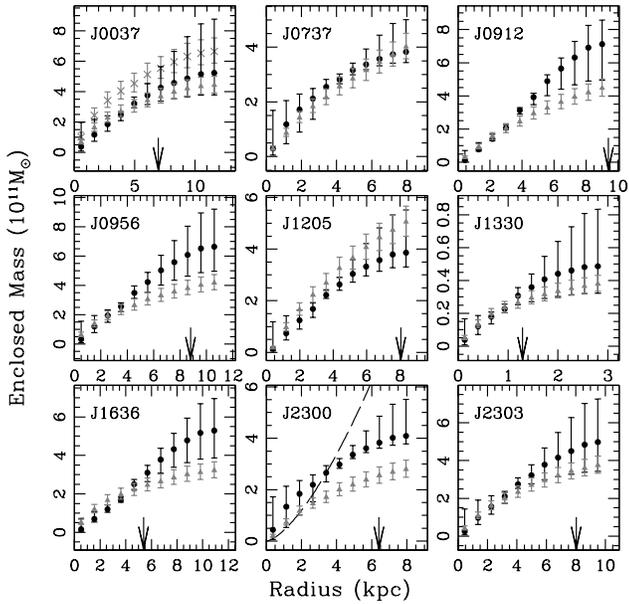}
\caption{Cumulative mass profiles of the lensing galaxies.  Black
circles and grey triangles denote total and stellar mass,
respectively. The error bars show the 90\% confidence region. The
effective radius ($\Re$; measured from the optical image) is given as
an arrow for each galaxy.  The enclosed mass is best constrained in
the region of the images, with the error bars enlarging at smaller and
larger radii.  Outside of $\Re$ all the galaxies have a significant
contribution from dark matter although some systems, most notably
J2300, present a very high dark matter contribution. A Chabrier IMF is
used for the stellar mass estimates.  The grey crosses on the upper
left panel correspond to the stellar mass for a Salpeter IMF, which
gives an unphysical $\Sigtot<\Sigstel$ in the inner parts of the galaxy.
The dashed curve in the bottom-middle panel shows $R^{1.5}$, the
steepest enclosed-mass profile (equivalently shallowest density
profile) allowed in our models.  The normalization of the dashed curve
is arbitrary.}
\label{fig:profile}
\end{center}
\end{figure}

With the methodology described in the previous two sections, we
generated non-para\-metric 2D maps of the total and stellar surface
density on the same pixelation.  Figure~\ref{fig:mass} shows both
distributions, the gray scale being $\Sigstel$ derived from population
synthesis and the contour maps indicating the $\Sigtot$ derived from
lensing.  (Both maps have uncertainties, shown in later figures, but
not here.)  Seven of the lenses have the inversion-symmetry constraint
mentioned in Section~\ref{sec:totmass}.
%; that is to say, they are
%symmetric with respect to rotation by $180^\circ<$.  
For the two lenses
where the images are well distributed in position-angle, J0912 and
J1636, we did not impose inversion symmetry and the models are allowed
to be lopsided.  The asymmetry in these two lenses is at the level of
10--15\% which given the uncertainties shown later in
Figure~\ref{fig:profile} is probably not significant.

We noted in Section~\ref{sec:totmass} that a model fitted to
point-like features can be used to predict extended images to some
degree.  For one case this can be achieved with a simple piece of
computer graphics.  If one draws the arrival-time contours of the {\em
point\/} source with a very close contour spacing, the resulting
pattern models the lensed image of an {\em extended\/} source with a
conical light profile \citep{2001AJ....122..585S}.
Figure~\ref{fig:J1205} illustrates for one lens, J1205+491, which
occupies the central panel of Figure~\ref{fig:mass}.  The white curves
are the contour lines, and two minima and two saddle points are
discernible; these are the locations of the point-like images.  In
most of the figure the contour lines are so close that the resulting
pattern is almost completely white, and has been airbrushed out.  In
the region shown, however, the white-on-black pattern closely
resembles the observed arcs.  The white lines have width $w=0.01\;\rm
arcsec$ on the scale of the figure.  The contour spacing in scaled
units \citep[see e.g., Equation 2.3 in][]{1986ApJ...310..568B} has
been chosen as $\Delta\tau=0.0015\;\rm arcsec^2$.  The implied source
radius in this type of model is $\Delta\tau/w$ or $0.15''$.
\cite{2006ApJ...649..599K} have much more detailed source maps (but
much simpler lens models) and this source size is typical.  The main
conclusion from Figure~\ref{fig:J1205} is that most of the lensing
information in the extended images is already present in the
point-like features.

Returning now to examine Figure~\ref{fig:mass} again, comparing the
stellar and total mass profiles suggests that
\begin{enumerate}
\item $\Sigstel$ and $\Sigtot$ tend to have aligned ellipticity, and
\item $\Sigstel$ falls off more steeply than $\Sigtot$.
\end{enumerate}

To test the statement (1) above, we compute the average ellipticity
and position angle of $\Sigstel$ and $\Sigtot$ using the second order
moments of the surface density.  Figure~\ref{fig:ellip} compares
these.  The orientations appear to be almost perfectly aligned.  The
tendency of lens models to be oriented with the light is well known
\citep[e.g.,][]{1997ApJ...482..604K}.  At the same time, the
magnitudes of the ellipticities is not correlated; in some galaxies
the stellar part is rounder, in some galaxies the dark matter is
rounder.

Statement (ii) above is addressed by Figure~\ref{fig:profile}, which
shows the circularly-averaged enclosed mass profiles, with 90\%
Bayesian confidence intervals.  The error bars in the total mass
derive from the ensemble of lens models, and have a characteristic
butterfly (or bow-tie) shaped envelope as explained in Section~2.
One envelope of the butterfly shape is expected to be approximately
$\Mtot(R)\propto R^{1.5}$, because the prior requires $\Sigtot$ to be
steeper than $R^{-0.5}$ (see~\S2).  This is indeed the case, as
illustrated in the bottom-middle panel of the figure. The other
envelope has a more complicated origin, having to do with the steepest
non-negative profile that can reproduce the observed image
positions. The
uncertainty in the stellar mass arises from the variety of
metallicities and star-formation histories compatible with the
available photometry (see Figure~\ref{fig:stellar}).  A
\cite{2003PASP..115..763C} form for the IMF is assumed, with the
effect of changing to a Salpeter IMF shown for one galaxy --- we
will return to this question below.

The strongest inference from Figure~\ref{fig:profile} is that dark
matter is located in halos.  While there is diversity in the details,
the general pattern is that stellar mass dominates in the inner
regions with the dark-matter fraction consistently increasing with
radius.\footnote{We remark that the above result cannot be an artefact
of the lens-modelling prior because (a)~the prior has no influence on
the normalization of the lensing mass, which depends only on image
positions, redshifts, and cosmological parameters, and (b)~a spurious
halo would amount to a $\Sigtot$ profile that was not steep enough,
whereas the prior imposes a {\em minimum\/} steepness.}  This
conclusion is as expected based on our knowledge of
stellar and gas dynamics in nearby galaxies, but is emerging here from
a completely different method and data set.

Our sample spans a similar range of masses and velocity dispersion
(see Table~\ref{tab:sample}). Except for J1330, the total mass extends
only over a factor of two.  All galaxies appear baryon dominated at
their centers, whereas outside the half-light radius the profiles show
a wide range of distributions.  Most of the galaxies show a
significant contribution from dark matter.  Interestingly, the only
galaxy with no evidence of dark matter to the radius observed (J0737)
is also the one where the observations are all interior of the
half-light radius.  This suggests that dark halos become significant
roughly around the half-light radius.

%%%%%%%%%%%%%%%%%%%%%%%%%%%%%%%%%%%%%%%%%%%%%%%
%%%%%%%%%%%%     FIGURE 5   %%%%%%%%%%%%%%%%%%%
%%%%%%%%%%%%%%%%%%%%%%%%%%%%%%%%%%%%%%%%%%%%%%%
%%%%%%%%%%%%%%%%%%%%%%%%%%%%%%%%%%%%%%%%%%%%%%%

%%%%%%%%%%%%%%%%%%%%%%%%%%%%%%%%%%%%%%%%%%%%%%%
%%%%%%%%%%%%     FIGURE 6   %%%%%%%%%%%%%%%%%%%
%%%%%%%%%%%%%%%%%%%%%%%%%%%%%%%%%%%%%%%%%%%%%%%
\begin{figure}
\begin{center}
\includegraphics[width=3.4in]{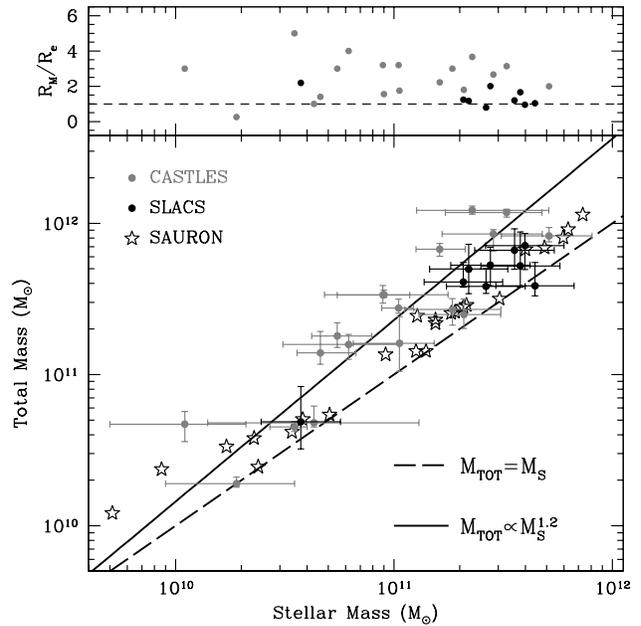}
\caption{Stellar and total mass compared in lensing galaxies and in
nearby galaxies with stellar dynamics.  {\em Main panel.} Dots show
the aperture masses $\Mstel,\Mtot$.  (The aperture is the region where
the total mass is usefully constrained by lensing.)  Black dots come
from the present work using SLACS lenses, and gray dots from a similar
analysis on a sample of CASTLES galaxies \citep{2005ApJ...623L...5F}.
The stars are measurements based on dynamical models applied to
(non-lensing) elliptical galaxies observed with the SAURON Integral
Field Unit \citep{2006MNRAS.366.1126C}. The dashed line corresponds to
$\Mtot=\Mstel$, while the dashed line corresponds to
$\Mtot\propto\Mstel^{1.2}$, which would give the observed tilt in the
Fundamental Plane.  {\em Top panel.} The aperture radius $\Rm$
relative to the half-light radius $\Re$.}
\label{fig:mtot}
\end{center}
\end{figure}
%%%%%%%%%%%%%%%%%%%%%%%%%%%%%%%%%%%%%%%%%%%%%%%

As mentioned above, we have assumed a Chabrier IMF for the stellar
population, and other realistic choices of IMF differ from this one by
a small factor not included in the error bars.  The simple power law
defined by \cite{1955ApJ...121..161S} and traditionally used as an
approximation to the IMF gives stellar masses $\sim 50\%$ higher.  For
some galaxies presented here, the implied stellar mass exceeds the
total mass as extracted from lensing. The upper-left panel of
figure~\ref{fig:profile} shows the stellar mass profile of J0037 for a
Salpeter IMF, which is clearly incompatible with the lensing results.
Hence, we can infer from our results that a Salpeter IMF is too
bottom-heavy. 

Our conclusion agrees with \cite{2006MNRAS.366.1126C}, who
used stellar dynamics to estimate the total mass in a sample of 25
E/S0 galaxies.

We can compare the aperture masses (stellar and total) with previous
work.  The aperture radius $\Rm$ we take as $2R_{\rm max}-R_{\rm
min}$, where $R_{\rm max}$ and $R_{\rm min}$ are the projected radii
of the outermost and innermost lenses images.  This $\Rm$ approximates
the radius within which lensing usefully constraints on the mass.  In
\cite{2005ApJ...623L...5F} the aperture radius was set for each galaxy
by seeing where the error bars became too large to be useful; such a
procedure tends to give slightly smaller aperture radii, but we will
disregard the difference here.  Figure~\ref{fig:mtot} shows the
aperture $\Mtot$ and $\Mstel$ in this work along with the earlier
lensing work \citep[labelled CASTLES:][]{2005ApJ...623L...5F} and the
stellar-dynamical work \citep[labelled
SAURON:][]{2006MNRAS.366.1126C}.  The galaxies from these three
independent data sets all show a clear trend towards a larger amount
of dark matter in more massive galaxies.  The observed tilt of the
Fundamental Plane suggests a scaling of $\Mtot\propto\Mstel^{1.2}$
\citep{2000MNRAS.316..786F,2003MNRAS.344..455F}, which appears consistent
with all the data.  However, the present sample is too small to test a
scaling law, especially because the range of aperture masses is small.
Also noticeable in the figure is that the SLACS galaxies tend to have
$R_M/\Re\simeq1$, unlike the CASTLES galaxies which cover a larger
range.  This is a selection bias due to the SLACS survey strategy.

In Table~\ref{tab:sample} we include an estimate of the dynamical
mass, using the popular relation $M_{\rm vir}=5\Re\sigma_e^2/G$
\citep[e.g.,][]{2006MNRAS.366.1126C}.  In the table $\Rm$ represents,
as explained above, the aperture radius usefully mapped by lensing.
Notice that for J0737 ($\Rm=0.8\Re$) the dynamical mass estimate is
much larger than the lensing or stellar masses within $R_M$, whereas
J1330 ($\Rm=2.2\Re$) is better mapped by the lensed images and give
compatible results between $\Mtot$ and the dynamical mass.

%%%%%%%%%%%%%%%%%%%%%%%%%%%%%%%%%%%%%%%%%%%%%%%
%%%%%%%%%%%%     FIGURE 7   %%%%%%%%%%%%%%%%%%%
%%%%%%%%%%%%%%%%%%%%%%%%%%%%%%%%%%%%%%%%%%%%%%%
\begin{figure*}
\begin{minipage}{14cm}
\includegraphics[width=6in]{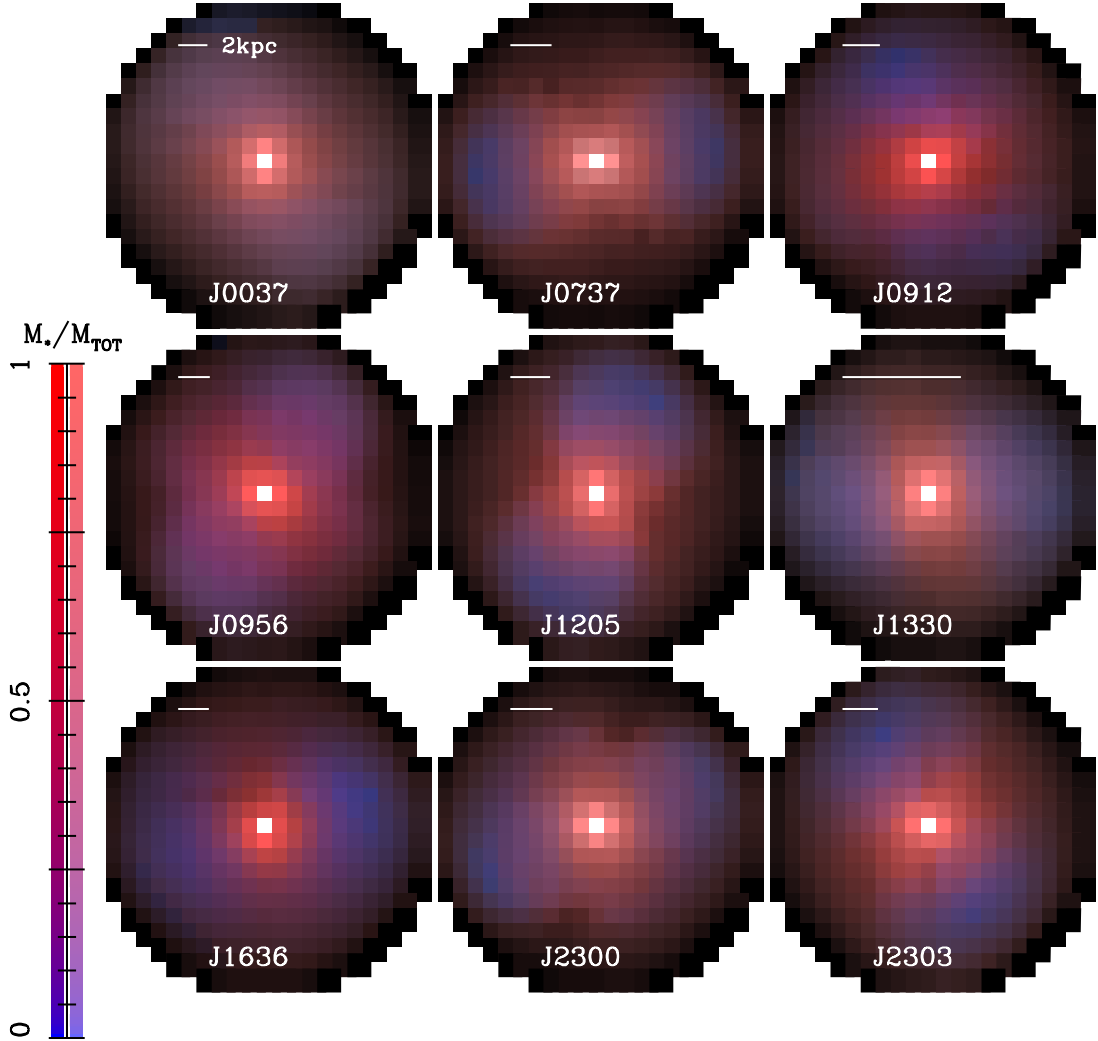}
\caption{A false colour map of stellar and total mass in galaxies.  The
surface mass density is show by intensity (black: low density; bright:
high density). Dark matter fraction is represented by colour (blue:
dark matter dominated; red: stellar matter dominated).  Uncertainty is
encoded by hue (pale: high uncertainty, bright: low uncertainty).  The
colour bar on the left shows a range of mass ratios for two different
uncertainties in the measurement of $\Sigstel/\Sigtot$: 0\% (exact
measurement; left) and 20\% error (right).  The horizontal bar in each
panel indicates 2~kpc.}
\label{fig:cmaps}
\end{minipage}
\end{figure*}
%%%%%%%%%%%%%%%%%%%%%%%%%%%%%%%%%%%%%%%%%%%%%%%

Finally, Figure~\ref{fig:cmaps} shows stellar and dark-matter maps in
false colour.  Here the brightness represents $\Sigstel$; the colour
represents the dark-matter fraction (red means all stellar, blue means
all dark); the hue represents the uncertainty (pale means more
uncertain).  The Appendix explains the colour-coding scheme more
precisely.  At the centers of the galaxies we see bright white,
meaning high density but with large uncertainty.  This turns into pale
red, indicating mainly stellar mass.  Further out we see faint pale
blue, indicating mainly dark matter.  This figure, though admittedly
ad hoc in its choice of colour coding, provides for the first time an
intuitive visual image of the distribution of dark matter in
individual galaxies.

\section{Conclusions}

The potential usefulness of early-type lensing galaxies for
understanding the interdependence of baryons and dark matter in galaxy
formation and evolution is widely appreciated. Lensing can be used to
map the total mass and starlight can be used to map the baryons.  In
this paper we do this for nine galaxies from the SLACS survey by
\cite{2006ApJ...638..703B}.  From the lensing data we derived
free-form pixelated models for the total mass, and from the galaxy
photometry we computed stellar population-synthesis models.  In both
casses we generated Monte-Carlo ensembles of models, in order to
marginalize over unknowns such as lensing degeneracies and
star-formation histories, thus obtaining realistic uncertainties. The
technique is basically the same as in \cite{2005ApJ...623L...5F}, but
whereas the earlier work only compared radial profiles now we compare
stellar and total mass in 2D.  Related work has been done on larger
samples of galaxies, but is limited to fitting simple parameterized
models for the lens, and does not model the stellar populations at all.

The 2D mass maps are shown in Figures \ref{fig:mass} and
\ref{fig:cmaps}.  The former overlays contours of $\Sigtot$ on a
grayscale of $\Sigstel$.  The latter shows the same information with
uncertainties as well, all encoded in false colour: red for stellar
mass, blue for dark, and pale versus coloured for uncertainty.  It is
evident that (a)~these galaxies are dominated by stars in the inner
regions, but mainly dark matter in the outer regions, and (b)~stellar
and dark components are well-aligned, but neither has a simple
elliptical shape.

One can of course still compute an ellipticity defined as a moment,
and this is shown in Figure~\ref{fig:ellip}. We see that ellipticities
of the stellar and total mass are uncorrelated in magnitude but are
almost perfectly aligned.  It would be interesting to see if this is
true of galaxy-formation simulations.

The profiles (Figure~\ref{fig:profile}) show that dark matter halos
begin to dominate around the half-light radius, although some galaxies
seem to present a stronger contribution from dark matter even inside
$\Re$ (e.g., J2300).  The present sample is too small to extract any
strong correlation of the dark matter distribution with global
properties such as total mass or luminosity. However, the trend found
previously \citep{2005ApJ...623L...5F}, namely that there should be
more dark matter in more massive galaxies, with a scaling of roughly
$\Mtot\propto\Mstel^{1.2}$ (which is equivalent to the tilt of the
Fundamental Plane) is compatible with the combined data of lensing
galaxies (labelled CASTLES and SLACS) along with the dynamical
analysis of local galaxies with the SAURON integral field unit
\citep{2006MNRAS.366.1126C}.

The top-left panel of Figure~\ref{fig:profile} also shows that a
Salpeter (1955) IMF cannot be used to estimate stellar masses as the
population synthesis models predict too much stellar mass compared to
the total mass obtained from our lensing studies. This result also
agrees with the analysis of the SAURON sample.

We emphasize that stellar and total masses are obtained from different
data through models of different physical processes.  There is no
tuning to give similar results.  That the stellar and total densities
come out compatible at the centers ---where the baryon content is
expected to dominate the mass budget--- indicates that systematic
effects are not significant.

\section*{Acknowledgements}
We thank Adam Bolton for many useful comments, and for making
available the surface-brightness fits to the lensing galaxies.

%%%%%%%%%%%%%%%%%%%%%%%%%%%%%%%%%%%%%%%%%%%%%%%%
%%%%%%%%%   REFERENCES   %%%%%%%%%%%%%%%%%%%%%%%
%%%%%%%%%%%%%%%%%%%%%%%%%%%%%%%%%%%%%%%%%%%%%%%%

%%%%%%%%%%%%%%%%%%%%%%%%%%%%%%%%%%%%%%%%%%%%%
\appendix

\section{False-colour maps}

For each galaxy we have sky-projected maps of three quantities: total
mass, uncertainty, and stellar-mass fraction.  How can we somehow
encode these into red, green, and blue, thus making a false-colour map?

In general, suppose we want to represent the following three
quantities:
\begin{eqnarray}
& & \hbox{$A$: an amplitude} \nonumber \\
& & \hbox{$f$: a ratio of components} \\
& & \hbox{$\Delta$: a fractional uncertainty} \nonumber
\end{eqnarray}
where $f$ and $\Delta$ both vary between 0 and 1.  A plausible mapping into
intensities $r,g,b$ of red, green, and blue is
\begin{eqnarray}
\label{rgb}
r/A &=& \athird\Delta + (1-\Delta) f \nonumber \\
g/A &=& \athird\Delta \\
b/A &=& \athird\Delta + (1-\Delta) (1-f). \nonumber
\end{eqnarray}
This gives total intensity $r+g+b=A$ regardless of $f$ and $\Delta$.
If $\Delta=0$, the colour will vary from blue at $f=0$ to magenta at
$f=0.5$ to red at $f=1$.  As the uncertainty $\Delta$ increases,
these colours will become paler, turning into white at $\Delta=1$.

The above scheme on its own is, however, not enough to produce useful
false-colour maps.  That is because the response of eyes to colour is
highly nonlinear, contrary to what Equation (\ref{rgb}) presupposes.
In practical image-processing, heuristic scale stretchings are always
necessary.  After some experimentation, we found the stretchings
\begin{equation}
A = \Sigtot^{1/2}, \qquad 
f = \left(\frac{\Sigstel}{\Sigtot}\right)^{5/2}, \qquad 
\Delta = \left(\frac{\Delta\Sigtot}{\Sigtot}\right)^{1/4}.
\end{equation}
to be useful. $\Sigstel$ and $\Sigtot$ represent the surface mass
density from the photometric and lensing analysis, respectively, in a
given pixel. $\Delta$ is the fractional uncertainty in the total
surface mass density, which dominates the error
budget. Figure~\ref{fig:cmaps} shows the results.

\end{document}